# The power of pictures: using ML assisted image generation to engage the crowd in complex socioscientific problems

Janet Rafner[a], Lotte Philipsen[a], Sebastian Risi[b], Joel Simon[c], Jacob Sherson[a,*]

[a]*Aarhus University, Denmark*
[b]*IT University of Copenhagen, Denmark*
[c]*Morphogen*

## Abstract
Human-computer image generation using Generative Adversarial Networks (GANs) is becoming a well-established methodology for casual entertainment and open artistic exploration. Here, we take the interaction a step further by weaving in carefully structured design elements to transform the activity of ML-assisted imaged generation into a catalyst for large-scale popular dialogue on complex socioscientific problems such as the United Nations Sustainable Development Goals (SDGs) and as a gateway for public participation in research.

## 1. Introduction

Artistic expression from both professional artists [31, 27, 32] and the general public [33, 35] is a key method for raising awareness of and facilitating discussions around the Sustainable Development Goals (SDGs). Generative Adversarial Networks GANs [7] are a natural tool to explore in this context, because they lend themselves to both professional and non-professional users in a variety of contexts. Professional artists use GANs as a new medium itself [8, 13, 14, 11], as a means of inspiring new ideas [8, 4, 15, 9, 10, 21] or as a helping hand for laborious tasks [6, 16, 17, 19, 20, 22]. By lowering the threshold for technical skills or capabilities, GANs also allow 'non-artists' or those with physical impediments to more easily express themselves creatively [1, 18, 2, 23]. Additionally, GAN images are genuinely phantom images – an AI model's imagination of what *images of the world* (not the world itself) could look like.

We believe that this technology not only holds the potential for casual entertainment and open artistic exploration but also, with a suitably designed interface, can act as a powerful engagement tool for people to visually communicate uncertainties that are difficult to express rationally, and to literally envision and imagine alternative ways of living. In this paper we adapted a novel GAN-based creativity assessment module: *crea.blender,* to facilitate discussions and agency towards solutions for the Sustainable Development goals (SDGs).

## 2. Existing project: Artbreeder

In recent years, platforms such as *Artbreeder* have successfully given the general public access to collaborative exploration of image generation [30]. *Artbreeder* (originally *Ganbreeder*) is a massive online tool for making images based on the idea of interactive latent variable evolution [3]. Images are 'bred' by selecting the generated offspring of one or more parent images in addition to direct 'gene' editing. *Artbreeder* operates as a hybrid of a tool and a social network, allowing users to share what they create and edit what they see. This community driven innovation allows certain images to go viral, spreading their 'DNA' throughout the image repository.

---

* Corresponding author at: sherson@phys.au.dk

## 3.   Existing project: crea.blender

As a recent adaptation of *Artbreeder*, *crea.blender* [12] supports quantitative investigation of creativity by letting players "blend" a restricted and carefully curated set of background free, hard coded source images into new images using the generator of BigGAN [24], which has been trained on ImageNet [25]. A pilot study found indications that *crea.blender* provides a playful experience, affords players a sense of control over the interface, and elicits different types of player behavior. Thus, more generally, the study indicated potential for the use of ML-assisted image generation for use in a scalable, playful, creativity assessment [12].

## 4.   New synthesis: Crea.blender: SDG edition

The stringent design of *crea.blender*, necessary to be in line with existing creativity research, also compromises some of the playful interactions of the original *Artbreeder* interface.

In an effort to regain elements of playful, open-ended exploration in a vast possibilities space we here present *crea.blender: SDG edition* (see figure) [36]. It retains the structured, goal- oriented setting of *crea.blender* with missions to create utopian and dystopian images of the future, but extends the open-ended creative potential by using instead the Landscape GAN of *Artbreeder* and allowing for free substitution of any of the six source images.

**Figure 1:** *crea.blender SDG edition* interface *Left)* participants are presented with 4 source images that can be 'swapped' with alternative images to suit the participants' liking. Participants playfully create new images by simply adjusting how much of each source image will be blended in. When the participants generate an image they are happy with, they can either save it as a 'utopian image' or a 'dystopian image' to convey their hopes or expectations of the future. *Right)* publicly available gallery of images from participants. Note: the image creation prompt can easily be adaptable for a particular context (e.g. life on earth).

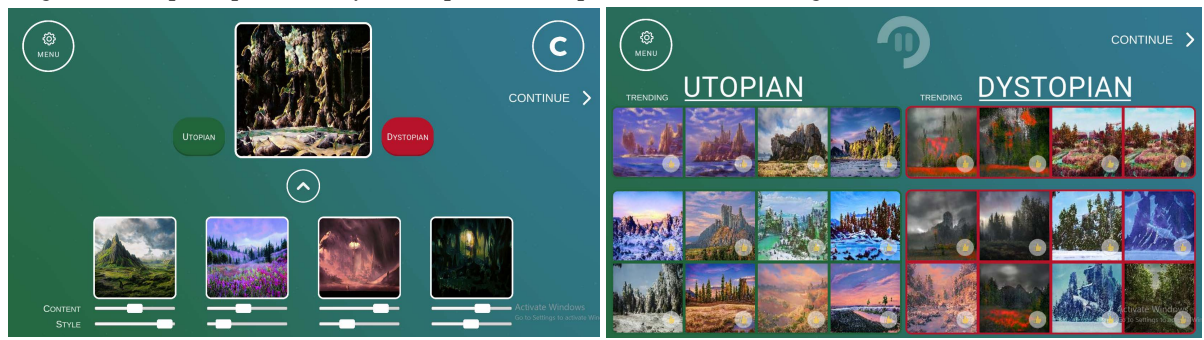

## 5.   Envisioned application areas

**SDG multi-stakeholder discussions:** Currently, *AI4good*, a leading action-oriented, global and inclusive United Nations platform [26], is in the process of curating a series on Artistic Intelligence. For this event the authors plan to launch *crea.blender: SDG edition*, as described above. The aim is for this game mode to lead to both physical and digital art exhibitions at major events such as Global Talent Summit, The World Economic Forum, and the G20 [28, 34, 29].

**Public participation in research:** In parallel we see the image generation interaction as a funnel of engagement, leading players to contribute to and engage in fundamental scientific questions. For example, players will be encouraged to play the full version of the *crea.blender* and the other parts of the game-based large-scale portfolio for creativity assessment where the participants can contribute to our ongoing research in creativity or play through a comprehensive behavioral economics/shared resources platform (both to be released on the authors' website).

In summary we demonstrate how cutting edge GAN interfaces are now at the tipping point of facilitating new means of imagination. They hold great possibility for catalyzing large-scale popular dialogue on complex socioscientific problems such as the SDGs and as a gateway for public participation in research, particularly if appropriate structured interfaces are employed.

**Ethical considerations**

While fully acknowledging the ethical problems of ImageNet [5] – especially in categories related to persons – we embrace the artificiality of GANs and believe that the creation of 'unreal' images are invaluable when remaking our real world. In the long term we hope to find alternative solutions to ImageNet, but in the meantime we do our best to mitigate these issues by using only landscape images.

Additionally, we are respectful of people's time and the participation in research and public debate that we are engaging them in. Thus, this interface will not stand alone, but will be paired with open, extensive information material on the scientific and civic engagement aspects, informed consent wherever there is login.

# 7. Appendix

Examples of utopian images created with crea.blender: SDG edition.

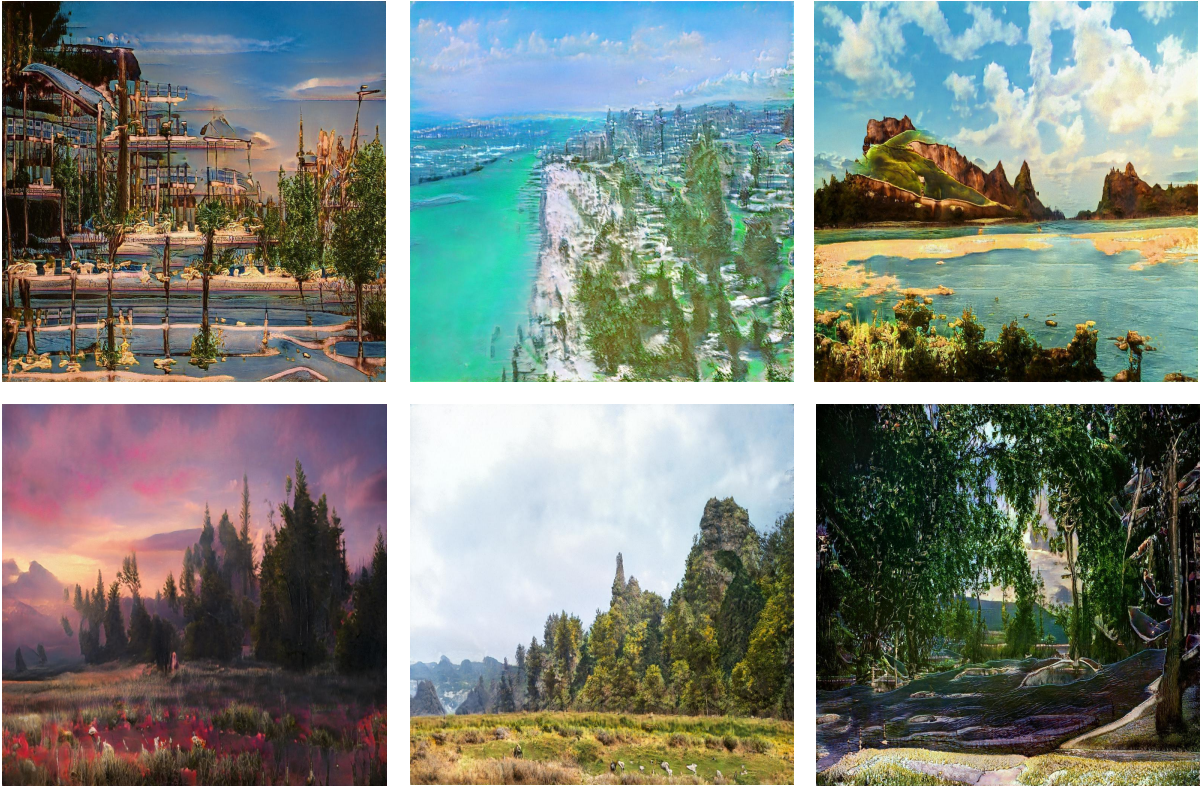

Examples of dystopian images created with crea.blender: SDG edition.

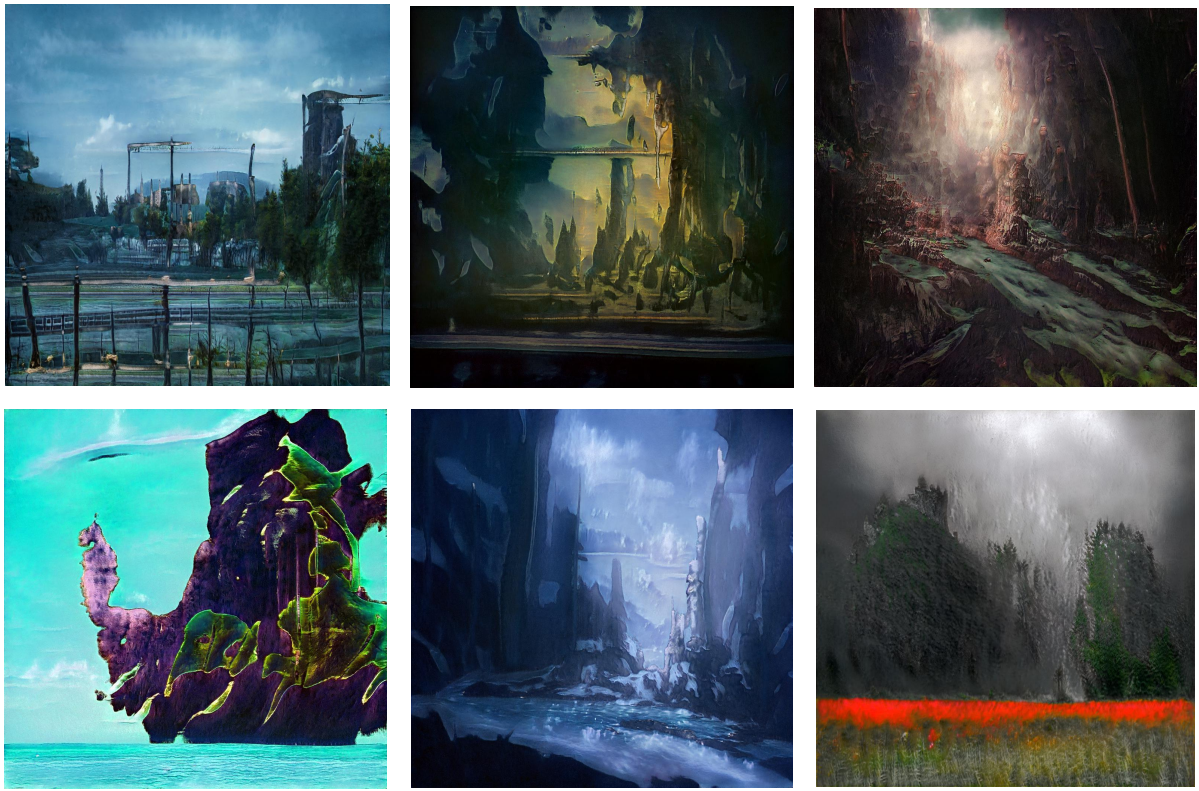